
\def\scalehalf{
  \font\tenrm=cmr10 scaled \magstephalf
  \font\tenbf=cmbx10 scaled \magstephalf
  \font\tenit=cmti10 scaled \magstephalf
  \font\tensl=cmsl10 scaled \magstephalf
  \font\tentt=cmtt10 scaled \magstephalf
  \font\tenex=cmex10 scaled \magstephalf
  \font\teni=cmmi10 scaled \magstephalf
  \font\tensl=cmsl10 scaled \magstephalf
  \font\tensy=cmsy10 scaled \magstephalf
  \font\seveni=cmmi8
  \font\sevenrm=cmr8
  \font\sevensy=cmsy8
  \font\fivei=cmmi6
  \font\fiverm=cmr6
  \font\fivesy=cmsy6
  \font\rm=cmr10 scaled \magstephalf
  \font\bf=cmbx10 scaled \magstephalf
  \font\it=cmti10 scaled \magstephalf
  \font\sl=cmsl10 scaled \magstephalf
  \font\tt=cmtt10 scaled \magstephalf
  \normalbaselineskip 13truept
  \mathfamilydefs}
\def\mathfamilydefs{
  \def\rm{\fam0 \tenrm}
  \def\it{\fam\itfam \tenit}
  \def\sl{\fam\slfam \tensl}
  \def\bf{\fam\bffam \tenbf}
  \def\tt{\fam\ttfam \tentt}
  \def\mit{\fam1 }
  \def\cal{\fam2 }
  \textfont0=\tenrm  \scriptfont0=\sevenrm  \scriptscriptfont0=\fiverm
  \textfont1=\teni  \scriptfont1=\seveni  \scriptscriptfont1=\fivei
  \textfont2=\tensy  \scriptfont2=\sevensy  \scriptscriptfont2=\fivesy
  \textfont\itfam=\tenit
  \textfont\slfam=\tensl
  \textfont\bffam=\tenbf \scriptfont\bffam=\sevenbf
       \scriptscriptfont\bffam=\fivebf
  \textfont\ttfam=\tentt
  \normalbaselines\tenrm}
  
\font \bigone=cmbx10 scaled\magstep1

\def\hang{\par\hangindent=\parindent\@}
\def\@{\noindent}
\def\cf{{\it cf}. }
\def\etal{{\it et~al}.}
%
%
\newcount\eqnumber
\eqnumber=1
\def\step#1{\global\advance#1 by 1}
\def\neweq{{\rm\the\eqnumber}\step{\eqnumber}}
\def\eqnew{\eqno(\neweq)}
{\scalehalf
\newcount\startpage
\startpage=1
\pageno=\startpage
\baselineskip=24pt plus .1pt
\voffset=-0.5truecm
\overfullrule=0pt
\tolerance=1500

\centerline{\bf{\bigone SHEAR FIELDS AND THE EVOLUTION OF
GALACTIC-SCALE DENSITY PEAKS}}
\bigskip
\centerline{\bf Rien van de Weygaert \& Arif Babul}
\medskip
\centerline{Canadian Institute for Theoretical Astrophysics}
\centerline{60 St. George Street, Toronto, Ontario M5S 1A7, Canada}
\bigskip
\centerline{ABSTRACT}
\medskip
We present preliminary results of our investigation into the
influence of shear fields on the evolution of galactic scale fluctuations
in a primordial Gaussian random density field. Specifically, we study
how the matter associated with a galaxy-scale peak evolves, to determine
whether the shear can affect the peak's ability to form a virialized structure.
We find that the evolution of the mass distribution in the
immediate surroundings of $\geq 1\sigma$ initial density
peaks is sensitive to the nature and magnitude of the shear.  Its final
fate and configuration spans a plethora of possibilities,
ranging from accretion onto the peak to complete disruption.
On the other hand, the mass defining the peak itself always tends to form
a virialized object, though a given galaxy-scale peak need not
necessarily form only one halo.  Furthermore, galaxy-size halos need not
necessarily be associated with initial galaxy-scale density peaks.
Under certain conditions, the shear field is capable of breaking up a
single primordial peak into two (and perhaps, more) distinct halos, or of
promoting the growth of smaller-scale peaks into galaxy-size halos.
\medskip
\@{\it Subject headings:} Cosmology: theory -- large-scale structure of the
Universe -- Galaxies: formation
\medskip
\@{\bf {\bigone 1. Introduction}}

It is generally believed that structure formation in the universe is the
result of initially small Gaussian random density fluctuations having
been gravitationally amplified, with objects such as galaxies having evolved
from peaks in the initial field. The statistical properties of peaks
in a Gaussian random field have been extensively studied (\cf
Peacock \& Heavens 1985, Bardeen \etal\ 1986). A major drawback in linking
the results of such studies to objects like galaxies is that there is no
clear association between features in the initial conditions and
eventual halos of a given mass. Consequently, halos of a characteristic
mass scale are generally identified, in a one-to-one mapping, with peaks in the
initial fluctuation field that is artificially filtered on the same scale.
This ignores the fact that actual collapse occurs where density and velocity
waves interfere constructively (Bond \& Meyers 1993). Apart from questions
regarding the validity of the identification procedure, smoothing itself
can lead to overinterpretations.
For example,
the mean density profile of a peak can be more extended than the filter
profile (Bardeen \etal\ 1986) and hence,
the actual mass associated with a peak can be considerably larger than the
mass scale of the filter.
Also, the filtering masks the complicated process by which structures form.

An important physical effect that gets relegated into the background by the
one-to-one peak-halo identification scheme is the influence of shear on
structure formation. Shear on an evolving density fluctuation can be due to its
own intrinsic asphericity (internal shear) or can be induced by the surrounding
mass distribution (external shear). Studies of collapsing homogeneous
ellipsoids have shown that internal shear can alter the collapse history of
structures (Icke 1973; White \& Silk  1979). Studies also show that
large-scale structures can induce significant tidal forces in their
vicinities ({\it e.g.} the Local Supercluster, see Lilje, Yahil \& Jones
1986) and that the resulting torques are capable of imparting angular
momentum to the protohalos (Efstathiou \& Jones 1979; Barnes \&
Efstathiou 1987) as well as, of influencing
the spatial and the kinematical structure of the virialized
halos (Binney \& Silk 1979; Dubinski 1992).
Hoffman (1986, 1989) was amongst the first to recognize the importance of
external shear on structure formation itself, noting that shear speeds up
collapse.
Bertschinger \& Jain (1993) and Bond \& Myers (1993) have recently
elaborated on various aspects of this. In particular, Bond \& Myers (1993)
find that tidal fields play an important role in the formation of low-mass
objects.

At this point, a clear distinction ought to be made between a collapse and
virialization. Collapse refers to a state where the local density
approaches infinity.  A positive energy structure that has evolved into a
thin expanding rod can be thought of as having collapsed, but not
virialized.
Collapse and virialization are synonymous only in the case of spherically
symmetric infall, although in the case of bound structures one will follow the
other fairly rapidly. In this study, we are interested
in the effect of cosmological shear fields on the
formation of virialized objects. Specifically, we generate a series of initial
density field realizations with a galaxy-scale peak at a designated location,
where we vary the height of the peak, its peculiar velocity, as well as the
magnitude and the orientation of the shear field acting upon it, and follow the
evolution of the associated matter distribution to see
whether the shear can cause the peak to spawn many
smaller clumps instead of forming a single halo. Such phenomena
may explain the results of Katz \etal\ (1993), who found that some of the
virialized objects in their N-body simulation did not correspond to
primordial density peaks. Also, we explore the
impact of shear on the fate of the matter distribution
immediately surrounding
the density peaks.
The degree to which this matter accretes onto
the halo determines its density distribution and final mass.

In this letter, we discuss the preliminary results of our study, concentrating
on four
cases.  In section 2, we describe our technique
for generating a constrained random initial density field.  In section 3, we
present our preliminary results, and in section 4, we attempt to draw some
conclusions.
\medskip
\@{\bf {\bigone 2. Initial Conditions}}

In order to generate constrained density fields (see Bertschinger 1987), we
use the prescription of Hoffman \& Ribak (1991).
They showed that in the
case where the constraints are linear functionals of the field, the problem
of generating a constrained random field has a simple and elegant solution.
The details of the method is best presented within the context of the problem
at hand (see Van de Weygaert \& Bertschinger 1993 for a more
comprehensive discussion).  Basically, we are interested in
studying the evolution of the matter distribution associated with
and in the neighbourhood of a galaxy-size peak in the initial
density field.  Apart from being able to determine the location, the scale
and the height of the peak, we also wish to sculpt the total matter
distribution in order to subject the peak to a desired
amount of net gravitational and tidal forces.  As a first step,
we need to define the term ``galaxy-scale peak''.  We identify such a peak as
a peak in the density field that has been smoothed by a Gaussian with a
characteristic scale $R_G= 0.585h^{-1}$ Mpc. (The enclosed mass is comparable
to that of an $L_*$-ish galaxy, $\approx 10^{12} M_{\odot}$.)

In addition to its scale and location, a peak in the smooth density field
is characterized by 18 constraints. We need to specify its height
and ensure that the 3 first derivatives of the smooth density field vanish at
its summit. The 6 second-order derivatives of the density field are
set by specifying the compactness, the axis ratios, and the
orientation of the peak. These 10 constraints
together determine the density distribution in the immediate vicinity of the
peak. The specification of the velocity field around the peak introduces
8 additional constraints: The 3 components of the smoothed
peculiar velocity field at the location of the peak
and the 5 independent components of the traceless shear tensor, which
when linearly extrapolated to the present are given by
$$\eqalign{{\bf v}_G({\bf r}_{pk})=&{\displaystyle H_0 {\cal F}(\Omega_0)
\over \displaystyle 4\pi}
\int d{\bf y}\int d{\bf x}\  \delta({\bf x}) W_G({\bf x},{\bf y})\,
{({\bf y}-{\bf r}_{pk}) \over\left\vert {\bf y} -{\bf r}_{pk}\right\vert^3}
\,,\cr
\sigma_{G,ij}({\bf r}_{pk})=&{\displaystyle 1 \over \displaystyle 2}
\left\{ {\displaystyle \partial v_{G,i} \over \displaystyle\partial r_j} +
{\displaystyle\partial v_{G,j} \over \displaystyle\partial r_i}\right\}
-{\displaystyle 1 \over \displaystyle 3} \left(\nabla \cdot {\bf v}_G\right)
\delta_{ij}\,\Biggl\vert_{{\bf r}={\bf r}_{pk}}\Biggr. ,\cr}\eqnew$$
\@where $\delta({\bf x})$ is the unsmoothed density
fluctuation field linearly extrapolated to the present,
$W_G({\bf x},{\bf y})$ is the Gaussian smoothing function,
$H_0$ is the Hubble constant and ${\cal F}(\Omega_0) \approx \Omega_0^{0.6}$.

Let us denote the constraints as a set of linear functionals $C_i[\delta]$ on
the constrained field $\delta$ with the value $c_i$ at the location of the
peak,
$\left\{ C_i[\delta;{\bf r}_{pk}] =c_i;\ \ i=1,\ldots,M \right\}$.
These linear constraints can be cast in the form of a convolution
between $\delta({\bf x})$ and some kernel
$H_i({\bf x};{\bf r}_{pk})$,
$$C_i[\delta;{\bf r}_{pk}]=
\int\,d{\bf x}\,f({\bf x}) H_i({\bf x},{\bf r}_{pk})=
\int\,{\displaystyle d{\bf k} \over \displaystyle (2\pi)^3}\,{\widehat\delta}
({\bf k}) {\widehat H}_i^{\ast}({\bf k})=c_i\,,\eqnew$$
\@where $\widehat\delta({\bf k})$ and ${\widehat H}_i({\bf k})$ are the Fourier
transforms of $\delta({\bf x})$ and $H_i({\bf x};{\bf r}_{pk})$, respectively.
The kernels for the peculiar velocity and the shear constraints, for example,
are
$${\widehat H}_{\bf v}({\bf k})=-i H_0 {\cal F}(\Omega_0) {\displaystyle
{\bf k} \over \displaystyle k^2} {\widehat W}({\bf k})e^{i {\bf k}\cdot
{\bf r}_{pk}}\,, \qquad
{\widehat H}_{\sigma_{ij}}({\bf k})=-H_0 {\cal F}(\Omega_0)\,
\left({\displaystyle k_i k_j \over \displaystyle k^2} -
{\displaystyle 1 \over \displaystyle 3} \delta_{ij}\right)\,
{\widehat W}({\bf k}) e^{i{\bf k}\cdot{\bf r}_{pk}}\,,\eqnew$$
\@where ${\widehat W}({\bf k})=\exp(-k^2 R_G^2/2)$ is the Fourier transform
of the Gaussian filter. Note that the position of the peak enters into the
kernels via the phase ${\bf k}\cdot{\bf r}_{pk}$.  Finally, the constrained
fluctuation field $\delta({\bf r})$ can be constructed according to (see
Van de Weygaert \& Bertschinger 1993)
$$\delta({\bf r})=\int\,{\displaystyle d{\bf k} \over
\displaystyle (2\pi)^3} \left[ {\widehat{\widetilde\delta}}({\bf k})+
P(k){\widehat H}_i({\bf k}) \xi_{ij}^{-1} (c_j-{\widetilde c}_j)\right]
e^{-i{\bf k}\cdot{\bf r}}\,,\eqnew$$
\@where $\widetilde\delta({\bf x})$ is an unconstrained
realization of the fluctuation field with the same power spectrum $P(k)$
as that characterizing the constrained field $\delta({\bf x})$,
${\widetilde c}_j$ is the value of the linear functional $C_j$ for the field
$\widetilde\delta({\bf x})$, and the random Gaussian variate
${\widehat{\widetilde\delta}}({\bf k})$ is its Fourier transform.  Also,
$$\xi_{ij}=\langle C_i C_j \rangle=\int\,{\displaystyle d{\bf k} \over
\displaystyle (2\pi)^3} {\widehat H}_i^{\ast}({\bf k}) {\widehat H}_j({\bf k})
P(k)\,.\eqnew$$
\medskip
\@{\bf {\bigone 3. Simulation and Results}}

For our simulations, we assume an $\Omega=1$ universe, a Hubble constant of
$H_0=50$ km$/$s$/$Mpc ($h=0.5$), and adopt the cold dark matter power spectrum
of Davis \etal\ (1985) --- normalized such that $\sigma(8h^{-1}\hbox{Mpc})=
1.0$ at $a=1$, the present epoch (Davis \& Peebles 1983). The fluctuation
field is sampled onto a $64^3$ grid in a periodic box of length
$25h^{-1}\hbox{Mpc}$, transformed into positions and growing mode
velocities of $64^3$ particles, and evolved using the P$^3$M N-body
code (with the force softening distance, $\epsilon=0.019h^{-1}$ Mpc) of
Bertschinger \& Gelb (1991).
The different simulation configurations are listed in Table 1 ---
the numerical values listed are linear extrapolations at $a=1$.

Our tailor-made galaxy-scale peaks either are at rest
or are streaming at $1000$ km$/$s, a velocity that is approximately equal
to the rms velocity ($\approx 884$ km$/$s) of galaxy-scale peaks.
The value of the expansion scalar, $\nabla\cdot{\bf v}_G$, associated
with the peak is $-197.5$ km$/$s$/$Mpc for 1$\sigma$ peaks and twice that
for $2\sigma$ peaks.  The shear tensor at the location of the peak is
orientated so that the off-diagonal terms
are zero.  For most configurations, the diagonal term with the largest
magnitude
is positive (dilation) while the other two are equal and negative
(contraction). In Runs 7--9, the situation is reversed.
Since we wish to induce non-negligible shear, the magnitude
of the largest element is chosen to be $100$--$150$ km$/$s$/$Mpc on the
scale of $0.585h^{-1}\,\hbox{Mpc}$, $1.5$--$2$ times the dispersion
($\approx 70$ km$/$s$/$Mpc) for the diagonal shear components (Bond 1987).

In the cases where the gravitational and the tidal forces on the peak are
primarily due to one large mass concentration (as in Runs 1 \& 2), the
evolution of the core--envelope material (we define the core as that material
which is initially within a comoving radius of $0.78h^{-1}$ Mpc of the density
peak and defines the peak, and we define the envelope as that material which is
within $1.56h^{-1}$ Mpc from the peak and constitutes its immediate
surroundings) is not interesting as it ends up merging into the massive halo.
The rest of the simulations, however, clearly illustrate the impact of shear
on the fate of the envelope material.  In the absence of shear, the envelope
would accrete onto the density peak.  Shear complicates the nature of secondary
infall and hence, affects both the final mass and the density profiles
of the halos (Gunn \& Gott 1972; Hoffman \& Shaham 1985; Bond \& Myers 1993).
An analytical study by Zaroubi \& Hoffman (1993) also suggests that the
evolution of
the envelope is sensitive to tidal influences. This effect of shear on the
envelope
material is best seen in Runs 3 \& 4.

The first row in Figure 1 shows the matter distribution in Run 3, at $a=0.5$.
The shear on the $1\sigma$ peak is induced
by a set of mass concentrations aligned along a filamentary
structure. The left panel presents the particle distribution in a $3h^{-1}$ Mpc
slice centred on the segment of the filament where the matter associated with
the initial peak is located.  The core material is unaffected by shear
and forms a small halo (circled).  In contrast,
as the peak streams towards the filament, the increasing shear stretches
the envelope and eventually tear most away (see centre panel). Note
that two clumps that are massive enough to be galaxies have
condensed out from the debris of the disrupted envelope.  Identifying the
initial density peaks with the particles closest to the positions of their
summit, we find that the above two ``galaxies'' do not
correspond to any initial $\geq 1\sigma$ galaxy-scale peaks.  The location
of all such peaks within a thicker $5h^{-1}$ Mpc slice are plotted in the
right panel.  This results conflicts with the underlying assumption of the
peak-halo identification scheme that galaxies only form around galaxy-scale
density peaks.

In Run 4, the $2\sigma$ galaxy-size peak starts is initially at rest and
embedded in a filament. The second row in Figure 1 displays the matter
distribution at $a=0.9$. The particle distribution in a $2h^{-1}$ Mpc
slice through the simulation volume is presented in the left panel. The centre
panel offers a detailed view of the region enclosed by the box, where
the matter associated with our peak ends up.  The distribution of the
core--envelope material (right panel) shows that the core is in
the process of merging with a larger halo. It was not significantly affected
by the shear.  However, a good fraction of the envelope has been
stripped away and captured by other halos.

In hindsight, the profound influence of shear on the envelope material is not
surprising.  But, can shear affect the core?  To explore this, we imposed
a considerable shear ($\left|\sigma_{xx} \right|=350$ km$/$s$/$Mpc) on a
$1\sigma$ galaxy-scale peak that is initially at rest (Run 6).  The particle
distribution at $a=0.8$ in a $1.24h^{-1}$ Mpc slice through the entire
simulation volume is shown in the left panel of the third row in Figure 1. The
centre panel shows a detailed, $3h^{-1}$ Mpc thick slice of the region enclosed
by the box.  The circled mass concentrations are the fragments of the core of
the original density peak.  The two core pieces, shown in isolation in the
right panel, collapse to form two distinct halos.  This is an example of a
case where a single initial peak forms more than one halo.  Only one of the
two halos contains the peak particle associated with the original density
peak;  the other halo contains no peak particle corresponding to an
initial $\geq 1\sigma$ galaxy-scale peak and
would have been missed under the peak-halo identification scheme.
The arrows in the centre panel locate all the
peak particles in a slice that extends beyond the
particle slice by $1h^{-1}$ Mpc on either side.

The shear configurations discussed thus
far involve dilation in one direction and contraction in the other two.  In Run
8, the situation is reversed and the $1\sigma$ galaxy-scale under consideration
is initially embedded in a plane.  It is also given an initial velocity
perpendicular to the plane.  The left panel of the last row in Figure 1 shows
the particle distribution at $a=0.8$ in a $1h^{-1}$ Mpc thick slice centred on
the plane.  The envelope of the peak is widely dispersed over this plane but
more interestingly, the core region fragments and forms two distinct halos,
although the shear is not unusually large.  A close-up of the region enclosed
by the box is shown in the centre and the right panels.  The projected velocity
vectors (centre panel) reveal that the two halos are more or less virialized.
As the arrow in the right panel indicates, only one of the halos contains
a peak particle associated with an initial $\geq 1\sigma$ galaxy-scale
peak.  We note that the evolution of a $2\sigma$ peak
under the same conditions is markedly different (Run 7).  Although the peak
initially starts to dissipate, mergers with nearby small clumps reverses the
trend and gives rise to a group-size halo containing most of the core material
as well as about half of the original envelope.
\medskip
\@{\bf {\bigone 4. Conclusions}}

Although preliminary, our study
into the role of shear on the formation of galaxy-scale halos has produced
some tantalizing results. We find that the tidal forces induced by large-scale
inhomogeneities can significantly affect the matter distribution in and around
a primordial density peak, with the extent of the influence depending on the
size of the density fluctuation, on the strength and the orientation of the
shear field, and on whether its source is localized or distributed:

[1] When the shear is primarily caused by a single mass concentration,
the galaxy-scale peak ends up merging with this object ({\it e.g.} Runs 1 \&
2).

[2] When the source of the shear is distributed ({\it e.g.} a filament
or a wall), the core of the peak tends to evolve undisrupted, while the
envelope is deformed and even dispersed ({\it e.g.} Runs 3--5).
The determination of halo properties,
like its final mass, based on the spherically symmetric secondary infall model
are likely to be in error.

[3] When the tidal forces are sufficiently strong, and favourably
oriented with respect to the peak's mass distribution and peculiar velocity,
the peak can be torn into at least two distinct pieces, with each eventually
forming a galaxy-scale halo ({\it e.g.} Runs 6 \& 8).  In this case,  it is
quite likely that not all of the halos will contain peak particles
demarcating initial galaxy-scale density peaks. These halos are potential
sites of galaxy formation though they would not have been flagged as such
under the peak-halo indentification scheme.  We note, however, that
we have not found a case of an initial galaxy-scale peak that did not form a
halo, contrary to the findings of Katz \etal\ (1993).

[4] In regions of high shear, galaxy-size halos can spring up from
primordial density peaks on scales smaller than $0.6h^{-1}$ Mpc, the commonly
adopted ``galaxy-scale'' ({\it e.g.} Run 3).  The convergence of matter
streams due to shear boosts the rate of accretion onto the small peaks.

The frequency of occurrence of the above effects, particularly [3] and [4],
is currently under study.  If generic, they suggest that identifying
galaxy-scale peaks in the initial field as
sites of galaxy formation, in a one-to-one mapping, is perhaps an inaccurate
oversimplification.
\medskip
\@{\bf {\bigone Acknowledgements.}}

\@We would like to thank Ed Bertschinger, Dick Bond and Bernard Jones for
stimulating
discussion and insightful advice.  We are also grateful to Ed Bertschinger
for providing us with the P$^3$M N-body code.
RvdW has been supported by an NSERC International Fellowship.

\medskip
\@{\bf {\bigone References}}
\parskip=0pt plus1pt

\hang Bardeen, J.M., Bond, J.R., Kaiser, N., Szalay, A.S., 1986, ApJ, 304, 15
\hang Barnes, J., Efstathiou, G., 1987, ApJ, 319, 575
\hang Bertschinger, E., 1987, ApJ, 323, L103
\hang Bertschinger, E., Gelb, J.M., 1991, Computers in Physics, Mar/Apr
1991, 164
\hang Bertschinger, E., Jain, B., 1993, preprint
\hang Binney, J., Silk, J., 1979, MNRAS 188, 273
\hang Bond, J.R., 1987, in Hinchliffe, I., ed., Proc. Theor. Workshop on
Cosmology and Particle Physics, Berkeley. World Scientific, p.22
\hang Bond, J.R., Myers, S.T., 1993, CITA preprint 93/27
\hang Davis, M., Efstathiou, G., Frenk, C.S., White, S.D.M., 1985, ApJ, 292,
371
\hang Davis, M., Peebles, P.J.E., 1983, ApJ, 267, 465
\hang Dubinski, J., 1992, ApJ 401, 441
\hang Efstathiou, G., Jones, B.J.T., 1979, MNRAS, 186, 133
\hang Gunn, J.E., Gott, J.R., 1972, ApJ, 176, 1
\hang Hoffman, Y., 1986, ApJ, 308, 493
\hang Hoffman, Y., 1989, ApJ, 340, 69
\hang Hoffman, Y., Ribak, E., 1991, ApJ, 380, L5
\hang Hoffman, Y., Shaham, J., 1985, ApJ, 297, 16
\hang Icke, V., 1973, A\&A, 27, 1
\hang Katz, N., Quinn, T., Gelb, J.M., 1993, MNRAS 265, 689
\hang Lilje, P.B., Yahil, A., Jones, B.J.T., 1986, ApJ, 307, 91
\hang Peacock, J.A., Heavens, A.F., 1985, MNRAS, 217, 805
\hang Van de Weygaert, R., Bertschinger, E., 1993, in prep.
\hang White, S.D.M., Silk, J., 1979, ApJ, 231, 1
\hang Zaroubi, S., Hoffman, Y., 1993, ApJ, 414, 20

\vfill\eject
\centerline{TABLE 1: List of Simulation Configurations}
\bigskip
\hrule\vskip0.1truecm\hrule
$$\vbox{ \tabskip 1em plus 2em minus 5em
\halign to\hsize{#\hfil&\hfil#\hfil&\hfil#\hfil&\hfil#\hfil&\hfil#\hfil&
\hfil#\hfil& \hfil#\hfil&\hfil#\hfil&\hfil#\hfil&\hfil#\hfil&
#\hfil\cr
Run & Peak & $v_{Gx}$ & $v_{Gy}$ & $v_{Gz}$ & $\nabla \cdot {\bf v}_G$ &
$\sigma_{G,xx}$ & $\sigma_{G,yy}$ & $\sigma_{G,zz}$&\multispan2\hfil Fate
of\hfil\cr
\ \ & Height & & & & & & & & \hfil Core & \hfil Envelope \cr
    &  $\sigma$  &  &  km/s & & km/s/Mpc &\multispan3\hfil km/s/Mpc \hfil
& & \cr
\noalign{\medskip\hrule\medskip}
1 & 2 & 1000 & 0 &    0 & -395.0 & 100 &  -50 &  -50 &\hfil Captured & \hfil
Captured \cr
2 & 2 & 1000 & 0 &    0 & -395.0 & -50 &  100 &  -50 &\hfil Captured & \hfil
Captured \cr
3 & 1 & 1000 & 0 &    0 & -197.5 & -50 &  100 &  -50 &\hfil Collapse & \hfil
Stripped \cr
4$^\dagger$ & 2 &    0 & 0 &    0 & -395.0 & 150 &  -75 &  -75&\hfil Captured &
\hfil Partly stripped \cr
5$^\dagger$ & 1 &    0 & 0 &    0 & -197.5 & 150 &  -75 &  -75 &\hfil Collapse
&
\hfil Stripped \cr
6$^\dagger$ & 1 &    0 & 0 &    0 & -197.5 & 350 & -175 & -175 &\hfil Fragments
&\hfil Stripped \cr
7$^\ddagger$ & 2 &    0 & 0 & 1000 & -395.0 &  75 &   75 & -150&\hfil Collapses
&
\hfil Partly stripped \cr
8$^\ddagger$ & 1 &    0 & 0 & 1000 & -197.5 &  75 &   75 & -150&\hfil Fragments
& \hfil Stripped \cr
9$^\ddagger$ & 1 &    0 & 0 &    0 & -197.5 &  75 &   75 & -150&\hfil Collapses
& \hfil Stripped \cr
\noalign{\medskip\hrule}
}}$$
\line{All numerical values are linear extrapolations evaluated at the present
epoch\hfill}
\line{$^\dagger$Initial conditions for Runs 4,5 \& 6 generated using same
random number
seed\hfill}
\line{$^\ddagger$Initial conditions for Runs 7,8 \& 9 generated using same
random number
seed\hfill}

\vfill\eject
\centerline{\bf {\bigone Figure Caption}}

\vskip 1.0truecm

\@{\bf FIGURE 1}
First row: The matter distribution at $a=0.5$ in Run 3.  The left panel shows
the particle distribution in a $3h^{-1}$Mpc slice centred on the core (circled)
formed by our initial galaxy-scale peak, while the distribution of the
core-envelope particles are plotted in the centre panel. The right panel shows
the location of all peak particles associated with initial $\geq 1\sigma$
galaxy-scale peaks in a thicker slice.
Second row: The matter distribution at $a=0.9$ in Run 4. The left panel
illustrates the particle distribution in a $2h^{-1}$Mpc thick slice though the
simulation volume.  The centre and the right panels offer a more detailed view
of the region enclosed by the box, with the right panel showing the
distribution of the core-envelope particles.
Third row: The matter distribution at $a=0.8$ in Run 6. The particle
distribution in a $1.24h^{-1}$Mpc slice is presented in the left panel. The
centre panel shows the particle distribution in a $3h^{-1}$Mpc slice of the
region defined by the box.  The two core fragments are circled, and the
corresponding particles are also plotted in the right panel.  The arrows in
the centre panel mark the location of all peak particles associated with
initial $\geq 1\sigma$ galaxy-scale peaks in a yet thicker slice.
Fourth row: The matter distribution at $a=0.8$ in Run 8.  The left panel
illustrates the particle distribution in a $1h^{-1}$Mpc thick slice.  The
centre and right panel show a close-up of the region enclosed by the box.
The projected velocity vectors of the particles are plotted in the centre
panel and the core particles are plotted in the right panel.  The arrow marks
the location of the only peak particle associated with initial $\geq 1\sigma$
galaxy-scale peaks in the vicinity.

\vfill\eject\end